\documentclass[12pt,preprint]{aastex}
\slugcomment{accepted for publication in Astrophysical Journal}
\shorttitle{Probing \ion{O}{6} Emission in Galaxy Halos}
\shortauthors{B. Otte et al.}

\received{2003 February 26}
\begin{document}
\title{Probing \ion{O}{6} Emission in the Halos of Edge-on Spiral Galaxies}
\author{B. Otte\altaffilmark{1}, E. M. Murphy\altaffilmark{2}, J. C.
Howk\altaffilmark{3}, Q. D. Wang\altaffilmark{4}, W. R. Oegerle\altaffilmark{5},
K. R. Sembach\altaffilmark{6}}
\altaffiltext{1}{Department of Physics and Astronomy, The Johns Hopkins
University, 3400 N. Charles St., Baltimore, MD 21218; otte@pha.jhu.edu}
\altaffiltext{2}{Department of Astronomy, University of Virginia,
Charlottesville, VA 22903; emurphy@virginia.edu}
\altaffiltext{3}{CASS, University of California-San Diego, 9500 Gilman Dr., La
Jolla, CA 92093; howk@ucsd.edu}
\altaffiltext{4}{Department of Astronomy, University of Massachusetts, Amherst,
MA 01003; wqd@astro.umass.edu}
\altaffiltext{5}{Laboratory for Astronomy and Solar Physics, NASA/GSFC,
Greenbelt, MD 20771; oegerle@uvo.gsfc.nasa.gov}
\altaffiltext{6}{Space Telescope Science Institute, 3700 San Martin Dr.,
Baltimore, MD 21218; sembach@stsci.edu}


\begin{abstract}
We have used the {\em Far Ultraviolet Spectroscopic Explorer} to search for
\ion{O}{6} $\lambda\lambda1031.926, 1037.617$ emission in the halos of the
edge-on spiral galaxies NGC\,4631 and NGC\,891. In NGC\,4631, we detected
\ion{O}{6} in emission toward a soft X-ray bubble above a region containing
numerous H$\alpha$ arcs and filaments. The line-of-sight component of the
motion of the \ion{O}{6} gas appears to match the underlying disk rotation.
The observed \ion{O}{6} luminosities can account for $0.2-2$\% of the total
energy input from supernovae (assuming a full \ion{O}{6} emitting halo) and
yield mass flux cooling rates between 0.48 and 2.8\,M$\sun$\,yr$^{-1}$ depending
on the model used in the derivations. Based on these findings, we believe it is
likely that we are seeing cooling, galactic fountain gas. No emission was
detected from the halo of NGC\,891, a galaxy in a direction with considerably
high foreground Galactic extinction.
\end{abstract}


\section{INTRODUCTION}

In a seminal work, \citet{spitz} predicted that the \objectname[]{Milky Way} is
surrounded by a tenuous, hot corona. Provided that the temperature of such gas
was below the escape temperature for a galaxy, he showed that the gas should
extend several kiloparsecs above the midplane. Spitzer's original arguments have
been extended to a more general theory of the dynamical evolution of the gaseous
disks of spiral galaxies. This so called ``galactic fountain'' model (Shapiro \&
Field 1976; Bregman 1980) postulates that the hot gas produced by multiple
overlapping supernovae would be buoyant in the thin, cold disks of gas in spiral
galaxies. As the hot gas rises from the thin disk into the halo, it cools,
perhaps forming condensations of neutral material which then fall back towards
the thin disk as high- or intermediate-velocity clouds.

The existence of such gas in the Milky Way is now well documented through
observations of strong \ion{O}{6} absorption (e.g. Savage et al. 2000) and X-ray
emission \citep{bur}. The 1032/1038\,{\AA} doublet transition of Li-like oxygen
(\ion{O}{6}) is an important tracer of hot, collisionally ionized material. At
temperatures of $\sim 3\times 10^5$\,K, where gas of near-solar metallicity
cools most efficiently, emission from the \ion{O}{6} doublet can be the primary
coolant. Hence, observations of \ion{O}{6} emission provide a direct probe of
the cooling rate of material near this temperature.  Furthermore, because the
cooling of gas at temperatures of a few times 10$^5$\,K is so much more
efficient than that of material at higher (or lower) temperatures, \ion{O}{6}
emission traces rapidly cooling material that may condense into clouds in the
galactic fountain scenario. Observations of \ion{O}{6} from this ``transition
temperature'' gas in the halos of galaxies provide fundamental information on
the circulation of material within spiral galaxies.

An \ion{O}{6} absorption line survey of the Milky Way \citep{sav03} showed that
the observed \ion{O}{6} was best explained by a patchy disk corotating with the
Galactic plane with signs of outflows. \ion{O}{6} observed in emission in the
Milky Way has been associated with gas in or around the Local Bubble (Welsh et
al. 2002 and references therein) and an outflow from the Perseus arm (Otte,
Dixon, \& Sankrit 2003). With our improving understanding of the \ion{O}{6}
distribution inside the Milky Way, it becomes necessary to search for \ion{O}{6}
emission in other galaxies as well in order to be able to distinguish between
peculiarities and common properties of galaxies. We analyzed spectra of two
galaxies (\objectname[]{NGC\,4631} and \objectname[]{NGC\,891}) taken by the
{\em Far Ultraviolet Spectroscopic Explorer (FUSE)}. Both galaxies are nearly
edge-on (i.e. allow clear distinction between disk and halo gas) and known for
strong extraplanar emission in radio continuum, optical lines, and soft X-rays
(SXR). Thus, they appear to be good candidates to search for evidence of
galactic fountains. While the spectra of NGC\,891 did not reveal any \ion{O}{6}
emission, we detected the \ion{O}{6} doublet in NGC\,4631. This is the first
detection of \ion{O}{6} in emission in a spiral galaxy other than the Milky Way
and its companions.

\section{OBSERVATIONS AND DATA REDUCTION}

\subsection{NGC 4631}

\begin{figure}
\plotone{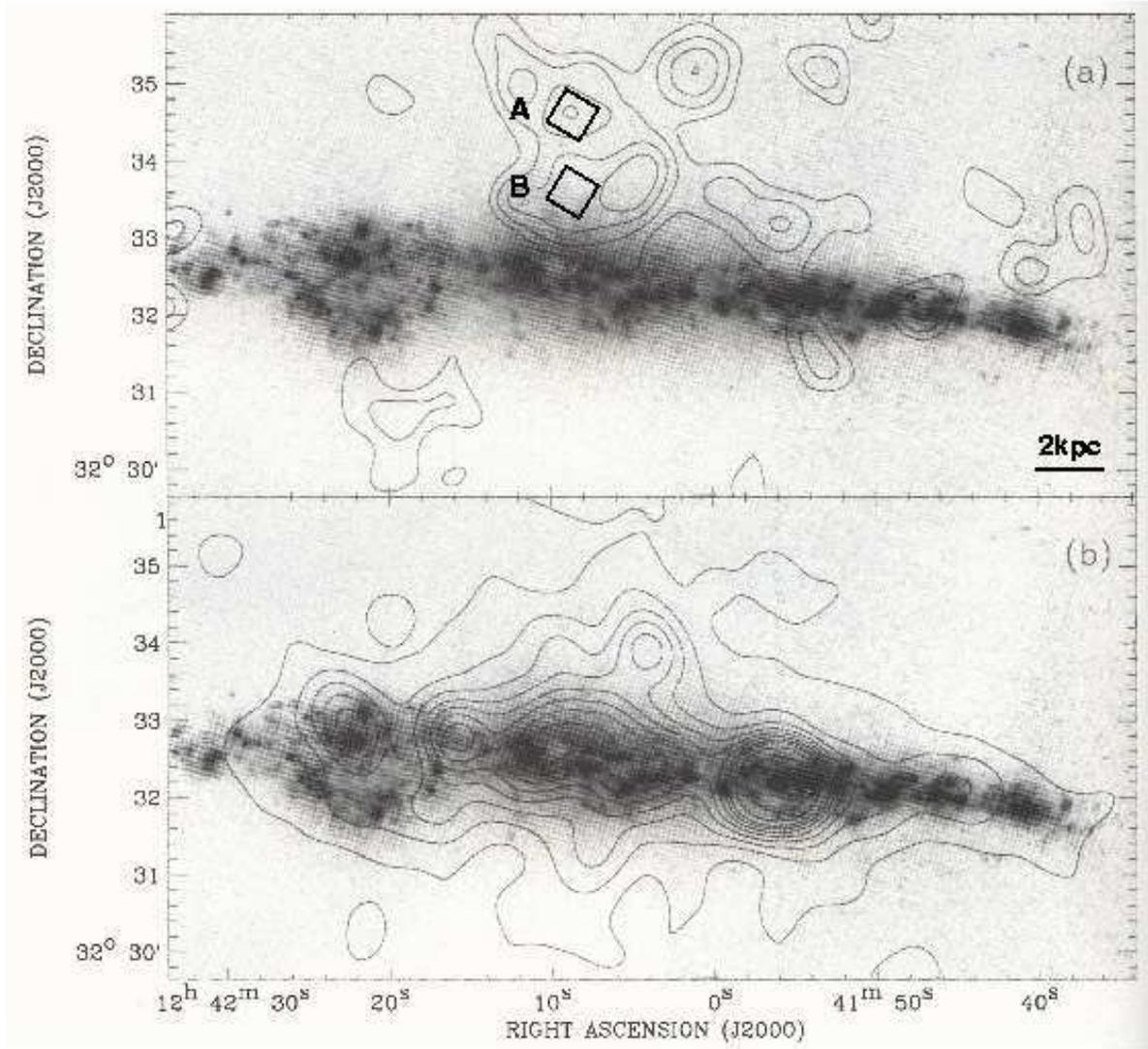}
\caption{\label{4631pos}X-ray contours superimposed on an H$\alpha$ image of
NGC\,4631 \citep{wang95}. Both {\em FUSE} positions are marked in the top panel,
which shows the SXR bubble observed in the {\em ROSAT} 0.15--0.3\,keV
band. The bottom panel shows the hard X-ray contours of the 0.5--2.0\,keV band
overlaid over the same H$\alpha$ image as in the top panel.}
\end{figure}

We obtained spectra with the low resolution aperture ($30\arcsec\times
30\arcsec$) of {\em FUSE} at two positions above the disk of NGC\,4631 (see
Figure \ref{4631pos}) as part of Project P134. Position A is at the peak of the
1/4 keV X-ray emission detected by \citet{wang95} using {\em ROSAT}. Position B
is between two H$\alpha$ arcs that were believed to form the walls of an open
chimney into the halo (Rand, Kulkarni, \& Hester 1992), although a recent {\em
Hubble Space Telescope} H$\alpha$ image \citep{wang01} indicates that this
feature may be a superposition of several loops. All {\em FUSE} observations
were conducted in TTAG mode where the detectors record the arrival time, x- and
y-position, and the pulse height of each event \citep{sahn}. About 2/3 of each
observation was conducted during orbital night.

The \ion{O}{6} lines in the NGC\,4631 spectra were extracted from the raw data
files. Only events with a pulse height between 4 and 15, inclusive, were used
in the analysis. Nearby \ion{O}{1} airglow lines were used to determine the
height of the extraction windows, and the spectra were summed perpendicular to
the dispersion direction. Emission free regions adjacent to the \ion{O}{6} lines
in the dispersion direction were used to determine the background.

\begin{figure}
\plotone{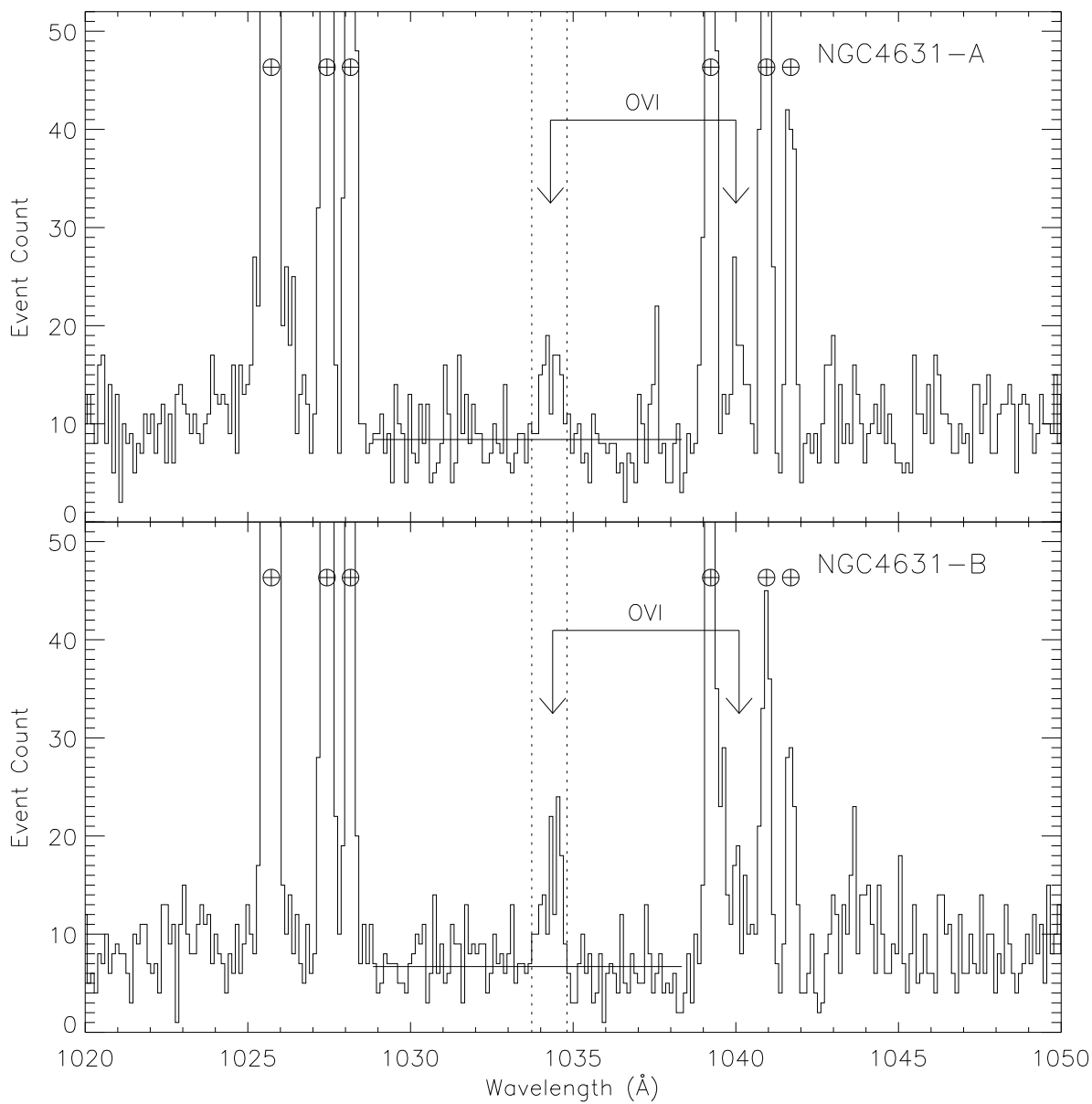}
\caption{\label{4631sp}{\em FUSE} day-plus-night spectra of NGC\,4631 at
Positions A (top) and B (bottom). The spectra were binned by 16 pixels for
display only. The \ion{O}{6}\,$\lambda\lambda$1032,1038 doublet is marked with
arrows; airglow lines are marked with the Earth symbol. The dashed lines show
the extraction window over which the \ion{O}{6}\,$\lambda$1032 photons were
counted. The continuum fit for this emission line is plotted as a horizontal
line. Its extend outside the dashed lines shows the region of the spectrum used
to determine the continuum.}
\end{figure}

Figure \ref{4631sp} shows a plot of the extracted spectra for both pointings
toward NGC\,4631. The wavelength scale was determined by running the spectra
through the {\em FUSE} data pipeline V2.0.5. Its uncertainty is about one pixel
(0.007\,\AA). Both the redshifted 1032\,{\AA} and 1038\,{\AA} line were detected
on the LiF1A channel. However, the latter was blended with the
\ion{O}{1}\,$\lambda\lambda$\,1039.23,1040.94,1041.69 airglow triplet. The
background at wavelengths $>1040$\,\AA\ was increased due to a stripe of
scattered light within the satellite. The \ion{O}{6} emission lines were also
seen on the LiF2B segment for Position B; however, the lower effective area
significantly reduced the quality of the data. Therefore, we have used only data
from the 1032\,{\AA} emission line of segment LiF1A in the analysis.

A comparison of day-plus-night versus night-only data yielded the same flux in
the \ion{O}{6}\,$\lambda$1032 line and therefore ruled out the possibility that
the emission arose from scattered solar \ion{O}{6} emission in the LiF channels.
Deep integrations towards other lines of sight ruled out detector artifacts,
scattered light, and stray light contamination at the positions of the detected
lines. The background was not significantly higher during the sunlit portion of
the orbit; therefore, we have used the combined day and night data in our
analysis. The resulting signal-to-noise ratios for the integrated
\ion{O}{6}\,$\lambda$1032 intensities are 4.5 (Position A) and 6.0 (Position B).

\subsection{NGC\,891}

\begin{figure}
\plotone{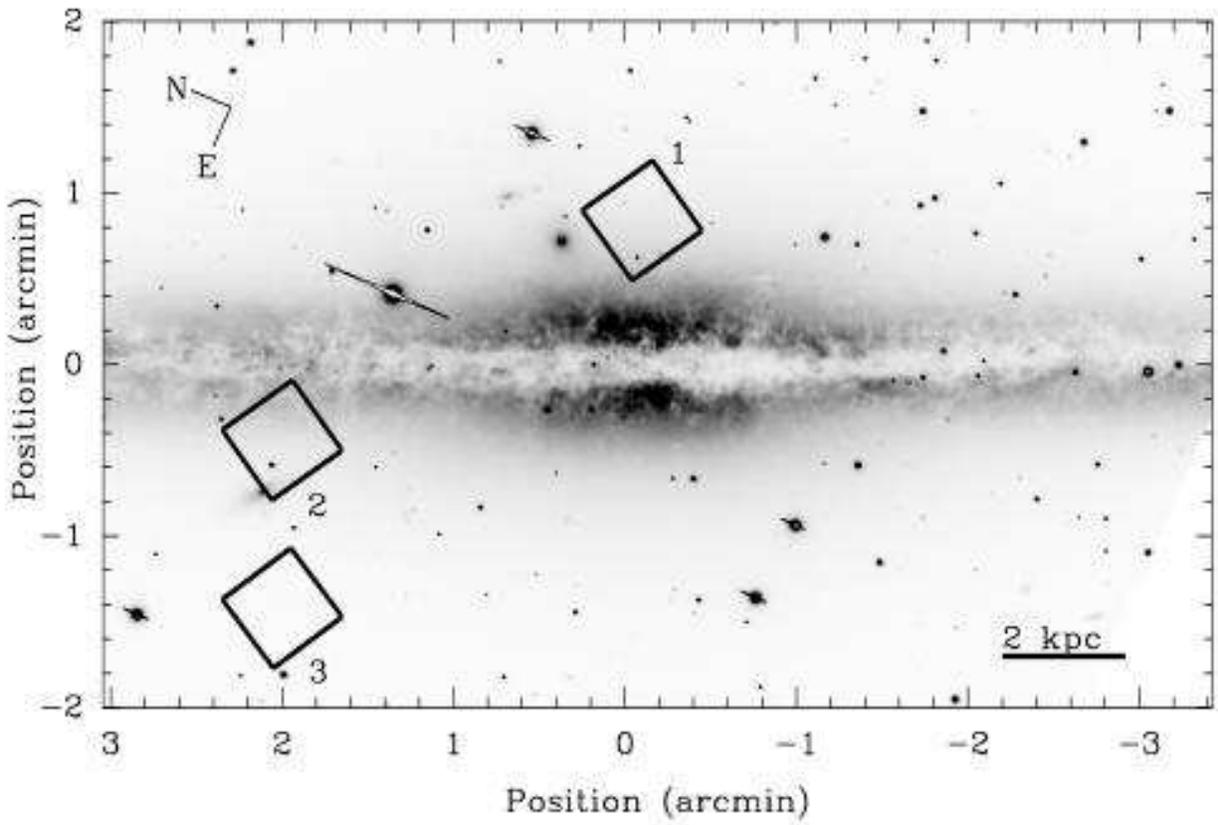}
\caption{\label{891pos}$BV$ image of NGC\,891 \citep{hs}. The three {\em FUSE}
positions are marked.}
\end{figure}

The data for NGC\,891 were obtained as part of {\em FUSE} Guest Investigator
Cycle 2 program B114. Two positions below the disk and one position above the
bulge of NGC\,891 (see Figure \ref{891pos}) were observed. The data reduction
was identical to the data reduction for NGC\,4631 above. Approximately 85\% of
the observations were conducted during the night portion of the orbit. The
observational data for both galaxies are listed in Table \ref{obs}. Given the
large random errors in the intensities of NGC\,891, we do not believe that any
of these detections are significant.

\section{RESULTS}

\subsection{\label{n4631}NGC\,4631}

NGC\,4631 is an Sc/SBd type galaxy at a distance of about 7.6\,Mpc with an
inclination of $i\approx85\degr$. The interaction between NGC\,4631 and its two
neighboring galaxies \objectname[]{NGC\,4656} and \objectname[]{NGC\,4627} might
have triggered the active star formation observed in NGC\,4631 today. The galaxy
possesses a complex system of H$\alpha$ filaments and bubbles \citep{wang01} and
an extended radio halo with spurs up to 10\,kpc above the disk \citep{hum}.
X-ray observations of NGC\,4631 revealed a SXR bubble on the north side of the
disk \citep{wang95} and a halo of gas extending up to 8\,kpc above the plane
with temperatures of $(2-7)\times 10^6$\,K \citep{wang01}. Both the radio and
the X-ray halo appear asymmetric with the larger extent being on the north side
of the disk.

The two {\em FUSE} pointings for NGC\,4631 were chosen to lie within the SXR
bubble observed with {\em ROSAT} (Wang et al. 1995, see Figure \ref{4631pos}).
Assuming an effective area of 27.0\,cm$^2$ for the LiF1A segment and
900\,arcs$^2$ for the aperture area, we derive \ion{O}{6}\,$\lambda$1032
intensities of $4600\pm1000$\,photons\,s$^{-1}$\,cm$^{-2}$\,sr$^{-1}$ (Position
A) and $8000\pm1300$\,photons\,s$^{-1}$\,cm$^{-2}$\,sr$^{-1}$ (Position B).
Table \ref{intensity} lists the intensities (also converted to
ergs\,s$^{-1}$\,cm$^{-2}$\,sr$^{-1}$) and the corresponding height above the
disk. The \ion{O}{6}\,$\lambda$1032 intensity at Position B is a factor of five
smaller than the 2\,$\sigma$ upper limit of \citet{ferg}, who observed the
southern portion of the galaxy with the {\em Hopkins Ultraviolet Telescope} on
the Astro-2 mission. The full extent of the \ion{O}{6} emitting gas is unknown.
However, it is useful to consider two cases. In the first scenario, the
\ion{O}{6} emission is confined to the region of the SXR bubble on the north
side of the disk. On the other hand, the \ion{O}{6} emission may extend across
the full hot halo revealed by the {\em Chandra} observations of \citet{wang01}.
The former would correspond to a single large bubble breaking out into the halo,
while the latter implies a full galactic fountain. Therefore, we will calculate
the luminosity for both cases assuming that the average \ion{O}{6} surface
brightness is equivalent to the average of Positions A and B. The
\ion{O}{6}\,$\lambda$1032/\ion{O}{6}\,$\lambda$1038 line ratio approaches two in
optically thin gas and is reduced in optically thick gas due to self-absorption.
We assumed the upper limit for this line ratio of 2 in order to account for the
\ion{O}{6}\,$\lambda$1038 emission in our calculations. Ignoring extinction (see
below) we calculated a lower limit for the total \ion{O}{6} luminosity of
$L_{\rm OVI}=(6.2\pm0.8)\times 10^{38}$\,ergs\,s$^{-1}$ for the bubble and
$L_{\rm OVI}=(2.3\pm0.3)\times 10^{39}$\,ergs\,s$^{-1}$ for the entire halo.

\subsubsection{\label{ext}Correction for Extinction}

The Milky Way extinction toward NGC\,4631 is $E(B-V)=0.017$ (Schlegel,
Finkbeiner, \& Davis 1998). Using a Galactic extinction curve by Cardelli,
Clayton, \& Mathis (1989), we derived an optical depth $\tau=0.25$, i.e. the UV
flux is attenuated by a factor of 1.28. The corrected values for the
luminosities are $L_{\rm OVI}=(7.9\pm1.1)\times 10^{38}$\,ergs\,s$^{-1}$ for the
bubble and $L_{\rm OVI}=(3.0\pm0.4)\times 10^{39}$\,ergs\,s$^{-1}$ for the
entire halo. The amount of extinction within NGC\,4631 is uncertain. Long slit
optical spectroscopy of the Balmer lines by \citet{mk} has revealed evidence for
significant and patchy extinction due to dust in the halo of NGC\,4631 up to
5\,kpc from the plane. However, another group found almost no extinction at a
height of $>1$\,kpc on the northern side of the disk \citep{otte}. We obtained
long slit spectra of NGC\,4631 with the 2.3\,m Steward Observatory on Kitt Peak,
Arizona, on 2002 April 7--9. The slit was about $4\farcs5$ wide and orientated
north-south to cross both {\em FUSE} positions. The measured H$\alpha$/H$\beta$
line ratios are comparable to those measured by \citet{otte}. We therefore
assume that extinction within the halo of NGC\,4631 is negligible.

\subsubsection{\label{sxr}Comparison with X-ray Luminosity}

As mentioned earlier, the \ion{O}{6} emission is an important coolant for hot
ionized gas. In fact, for gas in heating-cooling equilibrium, the \ion{O}{6}
emission cooling at intermediate temperatures (few $\times 10^5$\,K) should
dominate the radiative cooling in the SXR emission of $10^6$\,K gas \citep{sd}.
The SXR intensities in the $0.1-2$\,keV range measured by {\em Chandra} and
integrated over the area of the {\em FUSE} positions are $1.5\times
10^{-7}$\,ergs\,s$^{-1}$\,cm$^{-2}$\,sr$^{-1}$ (Position A) and $3.1\times
10^{-7}$\,ergs\,s$^{-1}$\,cm$^{-2}$\,sr$^{-1}$ (Position B) yielding
SXR/\ion{O}{6} ratios of $1.2\pm0.3$ and $1.3\pm0.2$ for Positions A and B, i.e.
the intensities are more or less comparable between the \ion{O}{6} doublet
and the SXR emission. However, the used \ion{O}{6}\,$\lambda\lambda$1032,1038
emission was a lower limit (see paragraph \ref{n4631}) and not corrected for
Galactic extinction (because we did not know the Galactic extinction for the
SXR emission); the SXR/\ion{O}{6} ratios did not account for any dust in
NGC\,4631 either. Thus, the quoted ratios are probably lower limits suggesting
that the gas is not in equilibrium.

The total \ion{O}{6} luminosity (corrected for Galactic extinction, see
paragraph \ref{ext}), if assumed to cover the whole halo of NGC\,4631, is by a
factor of 9 lower than the two component fit of the X-ray luminosity
($2.7\times 10^{40}$\,ergs\,s$^{-1}$, Wang et al. 1995, corrected for different
distances). The corresponding temperature of this X-ray luminosity was
approximately $6\times 10^{5}$\,K, i.e. a temperature about twice as high as
the one for the peak of the \ion{O}{6} cooling curve. This again implies that
the gas is not cooling efficiently in \ion{O}{6}, i.e. it is not in equilibrium.
However, the uncertainties here are higher than for the comparison above, as we
now have compared observational data with a model using different
simplifications (e.g. for the shape and extent of the emitting halo).

\citet{wang95} estimated that the mechanical energy input from supernovae is
about $(1.5-15)\times 10^{41}$\,ergs\,s$^{-1}$, i.e. the observed \ion{O}{6}
luminosity can account for at most $\sim2$\,\% of the total energy input from
supernovae. If the gas is in equilibrium, then the remaining energy input has to
be used to transport the gas from the disk into the halo against the
gravitational potential of the galaxy and against the pressure of the magnetic
field lines and is partially lost due to cooling of $\sim 10^6$\,K hot gas by
species other than oxygen. From the cooling flow models of \citet{edch}, we
calculate that the full halo case is equivalent to a mass flux of
2.8\,M$_\sun$\,yr$^{-1}$ for cooling at constant density and
1.8\,M$_\sun$\,yr$^{-1}$ for cooling at constant pressure. In case of a single
cooling bubble the corresponding mass fluxes are 0.74\,M$_\sun$\,yr$^{-1}$
(isochoric) and 0.48\,M$_\sun$\,yr$^{-1}$ (isobaric). A rough estimate for the
mass flux derived from {\em ROSAT} observations is $\sim 1$\,M$_\sun$\,yr$^{-1}$
\citep{wang95}. Despite the large uncertainties due to unknown extinction and
the differences in the cooling models, the data suggest that the \ion{O}{6}
emitting region is larger than the SXR bubble, but smaller (within a factor of
$\sim2$) than the X-ray halo observed with {\em Chandra}. This means that
outflows are more significant in the central region of the galaxy, as indicated
by the filamentary structure in the H$\alpha$ image \citep{wang01}. If, in fact,
there was strong, patchy extinction along our lines of sight, the actual
\ion{O}{6} luminosities could be much higher than the observed luminosities and
could account for a significant fraction of the input energy from supernovae. On
the other hand, if the \ion{O}{6} emission originated in the interface between
cold clouds moving through the hot coronal gas, the relationship between SXR and
\ion{O}{6} emission would become more complicated.

\subsubsection{Rotation Velocity}

We fitted the observed \ion{O}{6}\,$\lambda$1032 emission line by a convolution
of a tophat (106\,km\,s$^{-1}$ wide to represent the filled {\em FUSE}
$30\arcsec\times30\arcsec$ square aperture) and a Gaussian profile (to represent
the spectral resolution of {\em FUSE}). We used $\chi^2$ calculations to
determine the best central wavelengths and intrinsic velocity dispersions for
the emission lines at Positions A and B. In order to estimate 1\,$\sigma$
uncertainties for these parameters, we varied each parameter independently until
the corresponding minimum $\chi^2$ increased by 1 relative to the best fit
$\chi^2$. The derived values are $\lambda_{\rm c}=1034.30\pm0.09$\,\AA\ and
${\rm FWHM}=200\pm50$\,km\,s$^{-1}$ for Position A and $\lambda_{\rm
c}=1034.37\pm0.08$\,\AA\ and ${\rm FWHM}=170\pm50$\,km\,s$^{-1}$ for Position B.
The corresponding radial velocities (geocentric) are $v=690\pm30$\,km\,s$^{-1}$
(Position A) and $710\pm20$\,km\,s$^{-1}$ (Position B). An offset of $-9.1$ and
$-9.5$\,km\,s$^{-1}$, respectively, yields local standard of rest velocities. If
the \ion{O}{6} emitting gas does not fill the LWRS aperture and is not centered
in it, the maximum additional offset in the line centroid is 50\,km\,s$^{-1}$
(for a point source at the border of the aperture). The line broadening due to
the satellite's orbital motion is $<8$\,km\,s$^{-1}$, i.e. well within the
uncertainties of the FWHM.

The intrinsic FWHMs, if assumed to be thermal widths, correspond to temperatures
of $(1.4\pm0.7)\times 10^7$\,K (Position A) and $(1.0\pm0.6)\times 10^7$\,K
(Position B). These temperatures are an order of magnitude higher than the
hottest component of the SXR emission. This suggests that the line widths are
caused by turbulent motion or differential rotation rather than thermal motion.
From position-velocity diagrams \citep{rand94}, one can expect an offset of
about 50--100\,km\,s$^{-1}$ between the systemic velocity (606\,km\,s$^{-1}$)
and the velocity of gas corotating in the disk at the distances of the {\em
FUSE} positions projected onto the major axis. A combination of the size of the
low resolution aperture (30$\arcsec$ or about 1.1\,kpc) and differential
rotation yields a velocity dispersion of up to 200\,km\,s$^{-1}$ in the galactic
plane near the center. The \ion{O}{6} velocity centroids and FWHMs closely match
those of the \ion{H}{1} gas in the disk of NGC\,4631. Thus, the line-of-sight
component of the motion of the \ion{O}{6} emitting gas appears to follow the
rotation of the underlying disk gas. Due to the galaxy's high inclination, the
velocity of the \ion{O}{6} emitting gas perpendicular to the disk is unknown.
Although not conclusive, this strong relationship to the disk gas implies that
the \ion{O}{6} emitting material originated in the disk.

\subsubsection{Scale Height}

If the gas has an exponential distribution in the vertical direction above the
disk, we can calculate the density scale height of the gas from the two
observations at Positions A and B using:
\begin{equation}
h=\frac{-2(z_{\rm A}-z_{\rm B})}{\ln(\frac{I_{\rm A}}{I_{\rm B}})}
\label{scale_height}
\end{equation}
where $z_{\rm A}$ and $z_{\rm B}$ are the vertical heights of the projected
aperture above the plane of the galaxy, and $I_{\rm A}$ and $I_{\rm B}$ are the
measured specific intensities at Positions A and B, respectively. The 2 in
Equation \ref{scale_height} results from the fact that the intensity is
proportional to the square of the density. Using the intensities and $z$ heights
from Table \ref{intensity}, we find
\begin{equation}
h=8.3\pm4.1\,{\rm kpc}.
\end{equation}
The large uncertainty in the density scale height results from the large
uncertainties in the measured specific intensities at Positions A and B.
However, it appears that the scale height of \ion{O}{6} gas in NGC\,4631 is
significantly larger than the 2.3\,kpc scale height found by \citet{sav03} for
the Milky Way. Using the intensities from paragraph \ref{sxr}, we find a scale
height of about 6.6\,kpc for the SXR emission in NGC\,4631, whereas Hummel,
Beck, \& Dahlem (1991) reported a scale height of about 8\,kpc for the magnetic
field. Scale heights derived from optical emission lines range between 0.9 and
1.2\,kpc, not considering the roughly constant intensity at $z>3$\,kpc
\citep{mk}. Given the uncertainties, the scale heights of the \ion{O}{6} and SXR
emission and the magnetic field appear to be comparable.

\subsection{NGC\,891}

\begin{figure}
\plotone{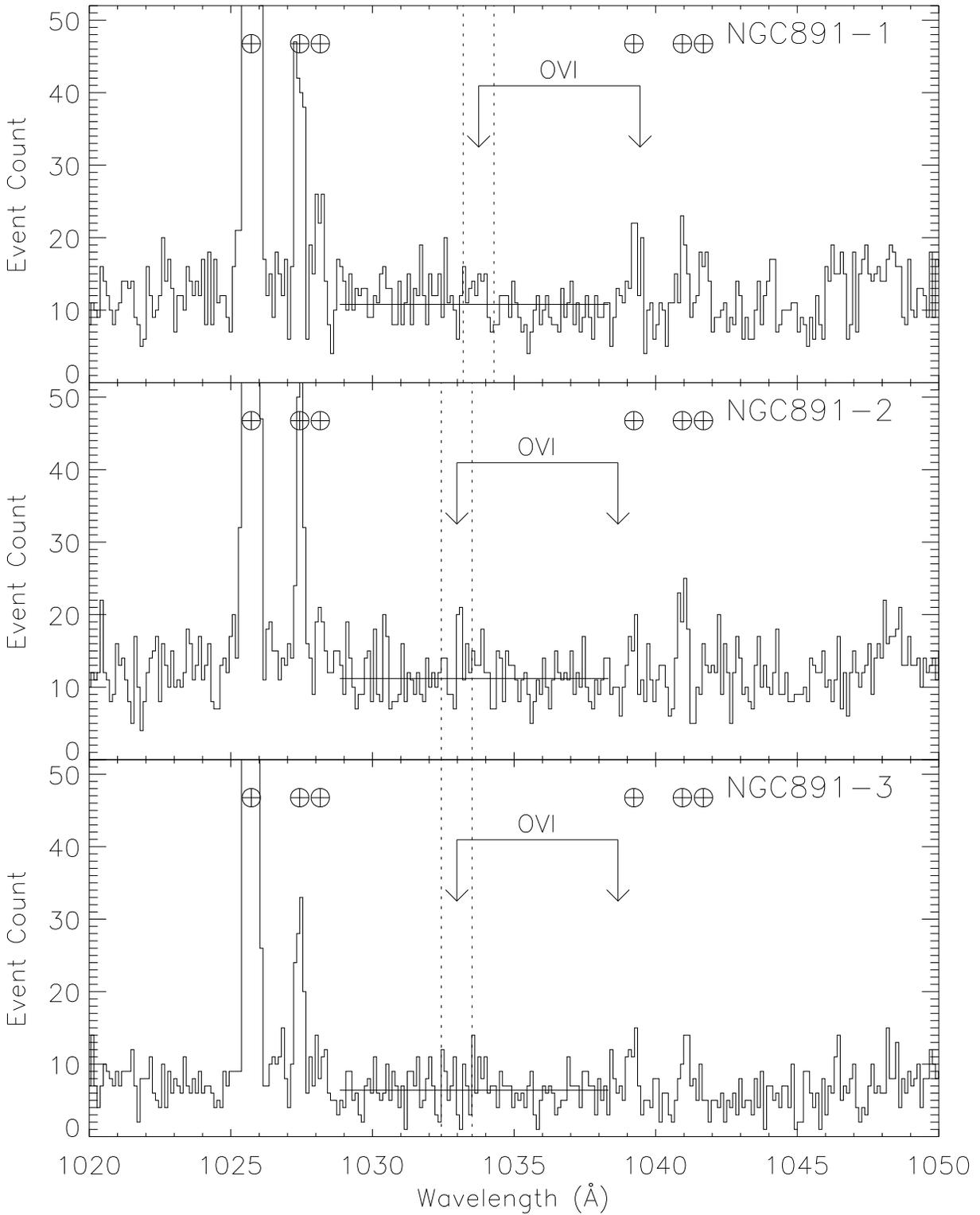}
\caption{\label{891sp}Same as Fig. \ref{4631sp}, but for NGC\,891 at Positions
1--3 (from top to bottom). The positions of the arrows and extraction windows
are based on the rotation curve by \citet{sa}.}
\end{figure}

The Sb spiral galaxy NGC\,891 is at a distance of about 9.6\,Mpc with an
inclination of $i\ge88.6\degr$ \citep{rup}. We observed three positions in the
halo of NGC\,891 and did not detect \ion{O}{6} in emission to a level greater
than $3\sigma$ (Fig. \ref{891sp}). The $3\sigma$ intensity limits can be
obtained from Table \ref{intensity}. The Milky Way extinction in this direction
is $E(B-V)=0.065$ \citep{sfd}. Assuming a standard interstellar extinction curve
\citep{ccm}, the UV flux is attenuated by a factor of 2.4 ($\tau=0.88$). If we
assume that the \ion{O}{6} emitting halo of NGC\,891 is the same size as the
X-ray emitting halo seen by \citet{bp}, and that the average specific intensity
of the halo in NGC\,891 is equal to the average at Positions 1 and 2, then the
upper limit to the \ion{O}{6}\,$\lambda$1032 luminosity is
$L_{1032}<7.5\times10^{38}$\,ergs\,s$^{-1}$. \citet{bh} found that the SXR
(0.1--2.5\,keV) band luminosity is $2.45\times10^{39}$\,ergs\,s$^{-1}$ with a
cooling rate of 0.08 M$_\sun$\,yr$^{-1}$ (corrected for $d=9.6$\,Mpc). This
predicts an \ion{O}{6}\,$\lambda$1032 luminosity \citep{edch} of
$5.7\times10^{37}$\,ergs\,s$^{-1}$ (isochoric) and
$8.8\times10^{37}$\,ergs\,s$^{-1}$ (isobaric), well within our limits. We have
not attempted any correction for extinction in the halo of NGC\,891 itself.

\section{CONCLUSIONS}

NGC\,4631 has a stronger radio halo than NGC\,891, but a comparable far-IR
luminosity. Rand, Kulkarni, \& Hester (1992) concluded from these observations
that NGC\,4631 probably has a larger star formation rate than NGC\,891. The {\em
Hubble Space Telescope} images of \citet{wang01} clearly show loops, arcs, and
filaments as would be expected from galactic chimneys. {\em ROSAT} observations
\citep{wang95} revealed the presence of a large region of SXR emission directly
above the arcs and filaments, indicating that some of the hot supernovae gas has
vented into the halo. In this paper, we reported the detection of \ion{O}{6}
emission at two locations above the arcs at the peaks of the SXR emission with a
line-of-sight velocity similar to the disk gas and a FWHM broader than what
might be expected from thermal motions of hot gas. The observation of \ion{O}{6}
emitting gas suggests that the hot gas originated in the disk and is cooling.
Due to the large uncertainties in the \ion{O}{6} intensities and luminosities
and the range in the luminosities and mass flux rates predicted by cooling flow
models, it is not clear whether the cooling flow in NGC\,4631 is (yet) in
equilibrium with the mass flux derived from X-ray observations. However, given
all this evidence, we believe that, because of its edge-on nature, NGC\,4631 is
the first spiral galaxy in which \ion{O}{6} has been detected in emission
unambiguously from the galactic halo and in which all the expected components
for a full cycle of gas in a galactic fountain have been observed. The lack of
reliable \ion{O}{6} measurements for NGC\,891 is likely explained with the large
foreground extinction of the Milky Way and the amount of dust in NGC\,891
itself. The \ion{O}{6} luminosities predicted from cooling flow models are well
within our 3$\sigma$ limits for NGC\,891.

\acknowledgements

The NGC 4631 data were obtained for the Guaranteed Time Team by the
NASA-CNES-CSA {\em FUSE} mission operated by the Johns Hopkins University.
Financial support to U. S. participants has been provided by NASA contract
NAS5-32985. The NGC 891 data were obtained as part of the {\em FUSE} Cycle 2
Guest Investigator Program B114. Financial support to U. S. participants has
been provided by NASA contract NAG5-10250. JCH recognizes support from NASA
grant NAG5-10957. This work made use of the NASA Extragalactic Database (NED)
and the NASA Astrophysics Data System (ADS).

\clearpage

\begin{deluxetable}{llcccc}
\tablewidth{0pt}
\tablecaption{\label{obs}OBSERVATIONS}
\tablehead{\colhead{Position} & \colhead{Program} & \colhead{Right Ascension} &
\colhead{Declination} & \colhead{Exposure Time} & \colhead{Events in LiF1A}\\
\colhead{} & \colhead{ID} & \colhead{(J2000 h m s)} & \colhead{(J2000 $\degr$
$\arcmin$ $\arcsec$)} & \colhead{(s)} & \colhead{1032\,\AA\ line}}
\startdata
NGC4631--A & P1340101 & 12 42 08.8 & 32 34 36.0 & 20954 & $55\pm 12$ \\
NGC4631--B & P1340201 & 12 42 08.8 & 32 33 36.0 & 16220 & $74\pm 12$ \\
NGC891--1 & B1140101 & 02 22 29.0 & 42 21 12.0 & 31603 & $15\pm 12$ \\
NGC891--2 & B1140201 & 02 22 40.0 & 42 22 36.0 & 31513& $25\pm 13$\\
NGC891--3 & B1140301 & 02 22 44.8 & 42 22 12.0 & 18531 & $12\pm 10$\\
\enddata
\end{deluxetable}

\begin{deluxetable}{lccc}
\tablewidth{0pt}
\tablecaption{\label{intensity}SPECIFIC \ion{O}{6}\,$\lambda$1032 INTENSITIES}
\tablehead{\colhead{Position} & \colhead{$z$} & \multicolumn{2}{c}{$I_{1032}$}
\\ \cline{3-4}
\colhead{} & \colhead{(kpc)} & \colhead{(photons\,s$^{-1}$\,cm$^{-2}$\,sr$^
{-1}$)} & \colhead{($10^{-8}$\,ergs\,s$^{-1}$\,cm$^{-2}$\,sr$^{-1}$)}}
\startdata
NGC4631--A & 4.8 & $4600\pm1000$ & $9\pm2$ \\
NGC4631--B & 2.5 & $8000\pm1300$ & $15\pm3$ \\
NGC891--1\tablenotemark{a} & 2.4 & $< 2000$ & $<3.8$ \\
NGC891--2\tablenotemark{a} & 1.4 & $< 2200$ & $<4.2$ \\
NGC891--3\tablenotemark{a} & 4.1 & $< 2800$ & $<5.5$ \\
\enddata
\tablenotetext{a}{The listed intensity is a $3\sigma$ upper limit.}
\end{deluxetable}

\clearpage

\clearpage

\clearpage

\clearpage

\end{document}